\documentclass[prl, footinbib, twocolumn, superscriptaddress]{revtex4-1}

\usepackage{amsmath}    
\usepackage{graphicx}   
\usepackage{color}      
\usepackage[colorlinks=true, citecolor=blue, linkcolor=blue]{hyperref}   

\usepackage{array,multirow}
\usepackage{pdfpages}

\usepackage{mdwlist} 
\usepackage{amssymb}
\usepackage[amssymb]{SIunits}    

\usepackage{bibunits}

\defaultbibliography{manuscript.bib}
\defaultbibliographystyle{apsrev4-1.bst}

\newcommand{\E}{\mathbb{E}}
\newcommand{\oneover}[1]{\frac{1}{#1}}

\newcommand{\dd}[2]{\frac{d #1}{d #2}}
\newcommand{\RNum}[1]{\uppercase\expandafter{\romannumeral #1\relax}}

\usepackage{etoolbox}
\makeatletter
\AtBeginEnvironment{thebibliography}{%
   \clubpenalty10000
   \@clubpenalty \clubpenalty
   \widowpenalty10000}
\makeatother

\begin{document}

\begin{bibunit}

\title{Two-Dimensional Clusters of Colloidal Spheres: Ground States,
  Excited States, and Structural Rearrangements}



\author{Rebecca W. Perry}
\affiliation{School of Engineering and Applied Sciences, Harvard
  University, Cambridge, Massachusetts 02138, USA}

\author{Miranda C. Holmes-Cerfon}
\affiliation{Courant Institute of Mathematical Sciences, New York University, 
New York, New York 10012, USA}

\author{Michael P. Brenner}
\affiliation{School of Engineering and Applied Sciences, Harvard
  University, Cambridge, Massachusetts 02138, USA}

\author{Vinothan N. Manoharan\textsuperscript{*,}}

\affiliation{School of Engineering and Applied Sciences, Harvard
  University, Cambridge, Massachusetts 02138, USA}
\affiliation{Department of Physics, Harvard University, Cambridge,
  Massachusetts 02138, USA}

\date{\today}

\begin{abstract}
  We study experimentally what is arguably the simplest yet
  non-trivial colloidal system: two-dimensional clusters of 6
  spherical particles bound by depletion interactions. These clusters
  have multiple, degenerate ground states whose equilibrium
  distribution is determined by entropic factors, principally the
  symmetry.  We observe the equilibrium rearrangements between ground
  states as well as all of the low-lying excited states. In contrast
  to the ground states, the excited states have soft modes and low
  symmetry, and their occupation probabilities depend on the size of
  the configuration space reached through internal degrees of freedom,
  as well as a single ``sticky parameter'' encapsulating the depth and
  curvature of the potential. Using a geometrical model that accounts
  for the entropy of the soft modes and the diffusion rates along
  them, we accurately reproduce the measured rearrangement rates.  The
  success of this model, which requires no fitting parameters or
  measurements of the potential, shows that the free-energy landscape of
  colloidal systems and the dynamics it governs can be understood
  geometrically.
\end{abstract}

\pacs{05.20.-y, 82.70.Dd, 02.40.-k, 05.10.Gg, 05.45.Tp}

\maketitle


Colloidal clusters containing a few particles bound together by weak
attractive interactions are among the simplest, non-trivial systems
for investigating collective phenomena in condensed matter.  Such
clusters can equilibrate on experimental time scales and display
complex dynamics, yet are small enough that the ground states can be
enumerated theoretically, and the positions and motions of all the
particles can be measured experimentally.  Theoretical and
experimental work on isolated three-dimensional (3D) colloidal
clusters of monodisperse particles has shown how the number of ground
states changes with the number of particles $N$~\cite{arkus09,
  arkus_deriving_2011, hoy_minimal_2010, hoy12,holmes14,hoy15} and how
the free energies of the rigid states are related to entropy-reducing
symmetry effects and entropy-enhancing vibrational modes~\cite{meng10,
  wales10, calvo12}.  The importance of entropy in colloidal clusters
stands in stark contrast to the case of atomic clusters, where
potential energy effects dominate. The entropically-favored clusters
are important clues to understanding nucleation barriers in bulk
colloidal fluids~\cite{crocker10, hoy12} and the local structure of
gels~\cite{royall_direct_2008}.

However, the excited states and structural rearrangements in such
clusters have not yet been studied experimentally.  In bulk materials,
local structural rearrangements are important to a variety of
dynamical phenomena, including the glass
transition~\cite{weeks_properties_2002},
aging~\cite{brito_heterogeneous_2007, yunker_irreversible_2009},
epitaxial growth~\cite{ganapathy10}, and the jamming
transition~\cite{lois_jamming_2008}.  A better understanding of the
internal dynamics in colloidal clusters could reveal local mechanisms
underpinning these bulk phenomena. Only a few experimental studies
have explored internal dynamics in colloidal clusters: Perry and
coworkers examined transitions between two states of a 3D 6-particle
cluster of spherical particles~\cite{perry12}; Yunker and coworkers
studied relations between the vibrational mode structure and the
contact network in disordered, two-dimensional (2D) clusters of
polydisperse particles as a function of $N$~\cite{yunker11,
  yunker_relationship_2013}; and Chen and coworkers examined the
interconversion and aggregation pathways in clusters of particles with
directional attractions~\cite{chen11a}.  As yet, however, a
quantitative understanding of the rearrangement rates and the pathways
through the excited states remains challenging.  Transition-state
models \cite{eyring35,wigner38,horiuti38,morgan14}, which relate
dynamics to the heights of saddle points on the energy landscape, are
not easily applied to colloids because the fluid surrounding the
particles damps and hydrodynamically couples their motions, and the
short-ranged interactions typical of colloidal particles are not
easily measured, making the topography of the landscape difficult to
accurately compute.  Indeed, as we shall show, the excited state
occupation probabilities and the transition rates are sensitive to
fine details of the potential, which are not easily measured.

\begin{figure}
    \includegraphics{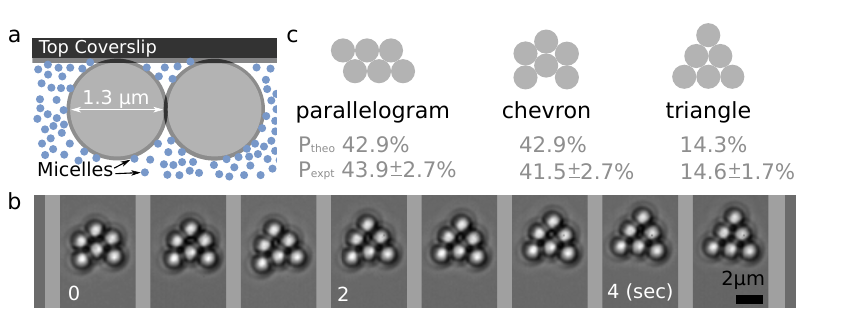}
    \caption{\label{fig:diagram} (a) SDS micelles induce a short-range
      depletion attraction between polystyrene microspheres and
      between the microspheres and the nearby glass coverslip. (b)
      Time-lapse images demonstrate a transition. (c) The three rigid
      ground states and their theoretical and experimental
      probabilities with 95\% confidence intervals~\cite{Note1} (the
      probabilities for the parallelogram include both chiral
      enantiomers).}
\end{figure}

We study experimentally the excited states and rearrangement rates in
perhaps the simplest type of colloidal cluster: isostatic arrangements
of equal-sized, spherical colloidal particles, constrained to lie on a
plane and held together by well-controlled,
short-range attractions a few times the thermal energy $k_BT$ in depth
(Figure~\ref{fig:diagram}a).  Because the clusters are isostatic, all
excited states have zero-frequency modes, or soft modes, in their
vibrational spectra (Figure~\ref{fig:diagram}b, \cite{Note1}).  By
tracking the particles over long times, we quantify the equilibrium
probability of each excited state and the motions of the particles
within each soft mode.  Surprisingly, the dynamics that emerge from
this landscape can be quantitatively described by a simple geometric
model involving only two parameters, a ``sticky parameter'' that
characterizes both the depth and curvature of the attraction, and a
diffusion coefficient, which we find to be insensitive to the mode.
Both parameters can be easily measured.  Therefore, no detailed
knowledge of the interactions or hydrodynamics is required to
reproduce the rates of rearrangement between ground states.

To make clusters, we first load an aqueous suspension of 1.3
$\mu$m-diameter sulfate polystyrene microspheres into a cell made from
two plasma-cleaned glass coverslips separated by 35 $\mu$m DuPont
Mylar\textsuperscript{\textregistered} A spacers~\cite{Note1}.  The
only additional component in the suspension is sodium dodecyl sulfate
(SDS), a surfactant that forms negatively charged micelles in
solution. The micelles create a weak depletion
interaction~\cite{asakura54, vrij76, iracki10} between the particles
and a stronger depletion interaction between the particles and
coverslip~\cite{lekkerkerker11, kaplan94b}, as illustrated in
Figure~\ref{fig:diagram}a.  At 33.4 mM SDS, we observe that 2D
clusters bound to a coverslip frequently transition between states but
rarely split apart or merge~\cite{Note1}.  At this concentration, the
sodium counterions from the surfactant reduce the Debye length to 2.85
nm, setting the effective hard-sphere depletion range of the micelles
to 30 nm, just 2.3\% of the particle
diameter~\cite{tulpar01,iracki10}.  As a result, the electrostatic and
depletion interactions between the particles are short-ranged. There
is likely also a short-range van der Waals attraction, which we
estimate tapers off to $k_BT$ when the particle surfaces are 145 nm
apart~\cite{israelachvili2011}.

At the beginning of the experiment, we assemble clusters at the top of
the sample cell using optical tweezers.  We then turn off the tweezers
and record digital micrographs for the remainder of the experiment.
The clusters, which would normally sediment, remain at the underside
of the upper coverslip, confirming the depletion attraction.  We use
particle tracking algorithms to locate the particles~\cite{crocker96},
link the locations into trajectories through time, and automatically
identify the cluster configurations~\cite{Note1}.

We focus on 6-particle clusters because this is the smallest system
with multiple ground states. Because these clusters are bound by
short-range interactions, the potential energy is proportional to the
number of contacts or ``bonds'' between particles.  The 6-particle
clusters adopt three ground states with nine bonds each
(Figure~\ref{fig:diagram}c): the parallelogram (which has two
enantiomers), chevron, and triangle.  In aggregate, the clusters
occupy the parallelogram and chevron states for equal amounts of time
but spend only one third as much time in the triangle state
(Figure~\ref{fig:diagram}c).  The measured occupation probabilities
agree with the expectation for a statistical mechanics ensemble in
equilibrium.  To calculate the probabilities, we assume that the
translational, rotational, and vibrational degrees of freedom are
independent, the vibrational modes are harmonic, and the translational
contributions and potential energy differ negligibly among the 3
states~\cite{Note1}.  As seen previously in 3D clusters, the
differences in occupation probabilities are primarily due to symmetry,
which enters into the rotational contribution~\cite{meng10,crocker10}.
\begin{figure*}
      \includegraphics[trim= 1 0 1 0, clip]{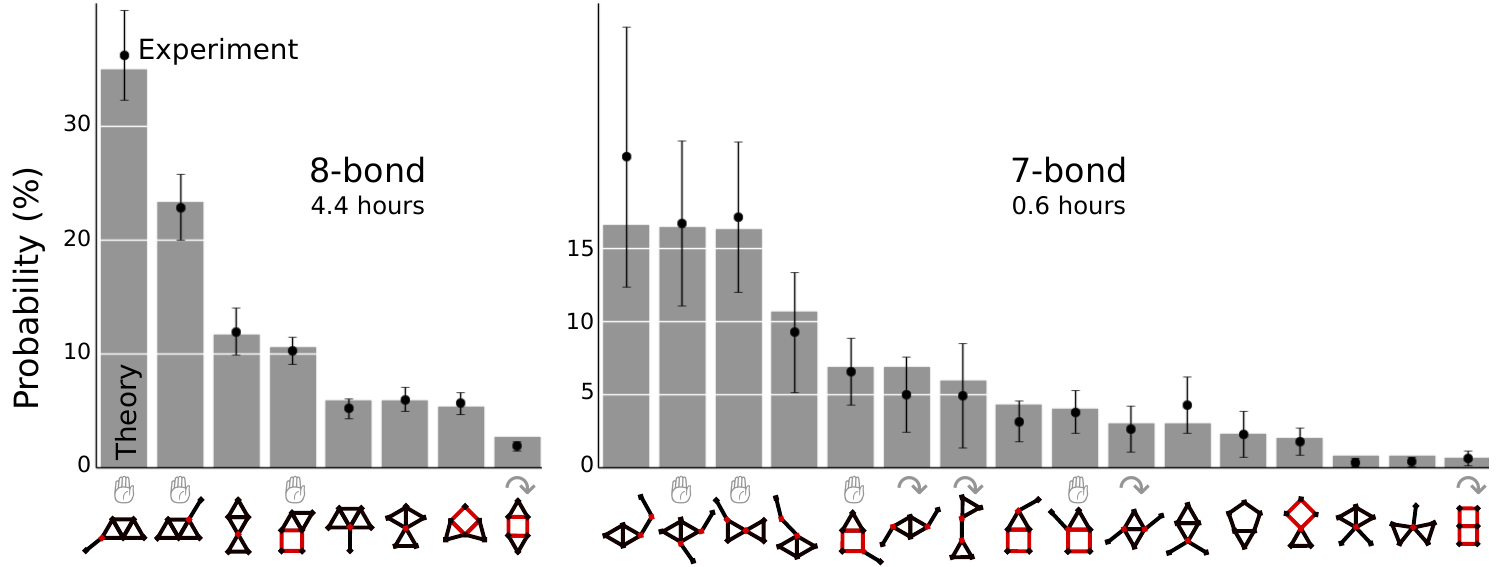}
      \caption{\label{fig:thryVexpt} Theoretical (bars) and
        experimental (points) probability distributions of the 8 and
        7-bond excited states.  Each bar-point pair is labeled by a
        connectivity diagram of the excited state, with hinge-like
        joints and non-rigid squares labeled in red.  Hand symbols
        mark the chiral states, and curved arrows mark the states with
        2-fold rotational symmetry (in 2D the only accessible symmetry
        axis is perpendicular to the plane of confinement).  The total
        observation time is 25.6 hours; for comparison, the clusters
        spend 19.5 hours in the 9-bond states.  Error bars represent
        95\% confidence intervals~\cite{Note1}.}
\end{figure*}

The excited states of the system have more complex and interesting
structures. All of them have zero-frequency modes.  The modes we see
at the 8-bond energy level have either hinge-like joints or
diamond-square-diamond~\cite{lipscomb66} flexibility
(Figure~\ref{fig:thryVexpt}).  Although the 7-bond energy level has
twice as many states, nearly all of the zero-frequency modes are
simply combinations of these two types of motion
(Figure~\ref{fig:thryVexpt}).  The exceptions are a state with a
flexible ring of five spheres and a state with a single sphere
detached from the cluster.  We do not include this disconnected state
in our 7-bond probability calculations because it is not a true
6-sphere cluster.

The fraction of time the clusters spend in the excited energy levels
depends on the surfactant concentration. At a concentration of 33.4 mM
SDS, the clusters spend 95.5\% of the time in states with 7 or more
bonds.  Of this time, 79.6\% is spent in ground states, 18.0\% in
8-bond excited states, and 2.4\% in 7-bond excited states.  As we
decrease the surfactant concentration, the distribution shifts toward
the excited energy levels.  Qualitatively, this shift makes sense,
since decreasing surfactant concentration corresponds to decreasing
depletion strength.  To understand the energy level occupation
probabilities quantitatively, we must consider the entropy of the soft
modes. We return to this point later.

Despite the wide variety of structures in the excited states, few
have any symmetry.  Surprisingly, the few symmetric states do not
occur as infrequently as we might expect, given the dominant role
symmetry---more specifically, permutational
entropy~\cite{meng10,gilson_symmetry_2010}---plays in the
probabilities of 6-sphere ground states in both 2D and 3D.
Furthermore, the asymmetric states have a highly non-uniform
distribution that is only partially explained by the increased
probability of states that are pairs of chiral enantiomers
(Figure~\ref{fig:thryVexpt}). These observations suggest that the
variation in probabilities arises from entropic factors other
than the permutational contribution.

\begin{table}
  \caption{Structural rearrangement rates between
    each of the ground states: (P)arallelogram, (C)hevron, and (T)riangle.  
    In total, we observed 820 transitions in 25.6 hours of data from 44
    clusters.  Measured values used in postfactor: $D=0.065$
    $\mu$m$^2$/s (234 $\mu$m$^2$/hr), $\kappa = 30.5$ and $d$ = 1.3
    $\mu$m.  
} 
\begin{ruledtabular}
\begin{tabular}{cccccccccc p{.45cm} c  cccc}
  &&\multicolumn{7}{ c }{Theory} &&&\multicolumn{3}{ c }{Experiment}&\\
  &&\multicolumn{3}{ c }{\scriptsize{(nondimensional)}} &&\multicolumn{3}{ c }{\scriptsize{(per hour)}} &&&\multicolumn{3}{ c }{\scriptsize{(per hour)}}&\\[-.5ex]
  &&\multicolumn{3}{ c }{\scriptsize{end state}} &&\multicolumn{3}{ c }{\scriptsize{}} &&&\multicolumn{3}{ c }{}&\\[-.5ex]
\parbox[t]{1mm}{\multirow{3}{*}{\rotatebox[origin=c]{90}{\scriptsize{ start state\ \ }}}} && P & C & T &&P & C & T&&&P & C & T\\

&\multicolumn{1}{r}{P}&1.17&1.43&0.67&&  5.3&6.5&3.0 &&& 4.4 & 5.5 & 2.5&\\

&\multicolumn{1}{r}{C}&1.43&2.31&0.56 &$\times\frac{D}{\kappa d^2}$=& 6.5&10.5&2.5 &&&  5.4 & 7.7 & 1.9\\

&\multicolumn{1}{r}{T}&0.67&0.56&0&&3.0 &2.5&0.00 &&& 2.5 & 2.2 & 0.04 

\end{tabular}
\end{ruledtabular}

\label{tab:transitions}
\end{table}

\begin{figure}
      \includegraphics[trim=0 1 0 0, clip, width=8cm]{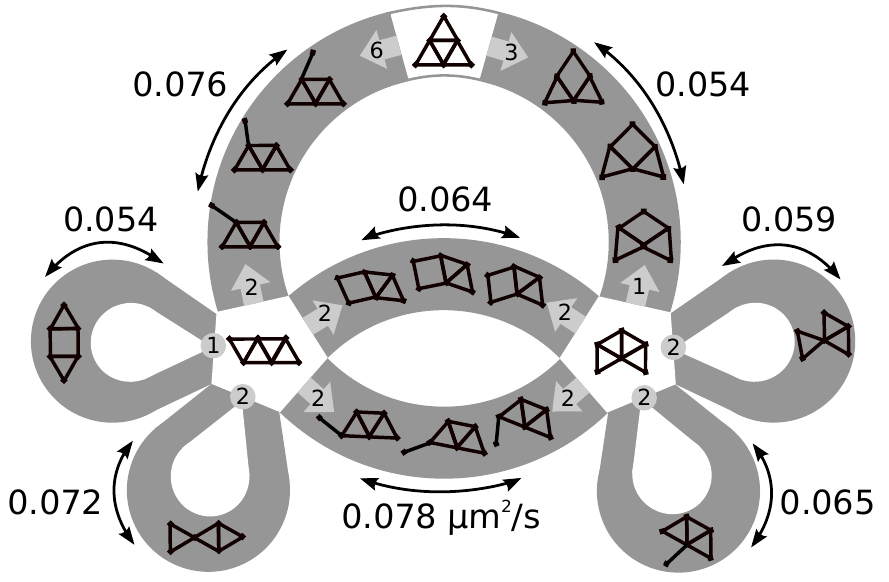}
      \caption{\label{fig:transitions} The 8-bond excited states form
        transition pathways (gray background) between the ground
        states (white background). The numbers at the
        edges of the ground states show the number of bonds that lead
        into each nearby pathway.  The measured diffusion coefficients
        of the modes range from 0.054 to 0.078 $\mu$m$^2$/s~\cite{Note1}.}
\end{figure}

We also measure the rate of rearrangements between ground states and
find that the matrix of rearrangements per unit time is symmetric
(Table \ref{tab:transitions}), as expected in equilibrium.  Most of
these rearrangements involve a single bond breaking, followed by the
cluster diffusing along the soft mode in its excited state and finally
forming a new bond to arrive at a ground state
(Figure~\ref{fig:transitions}).

Understanding the excited state probabilities and rearrangement rates
requires us to consider the entropy of the soft modes and the dynamics
along the resulting free-energy landscape.  In contrast to typical
molecular-scale transitions, in which the potential energy varies
along the entire reaction coordinate, our clusters first break out of
a narrow attractive well and then freely diffuse in soft modes at
constant potential energy under only an entropic driving force.  We
therefore expect the transition rates to depend on the entropy along
the modes, the hydrodynamic drag, and the distance to diffuse in the
soft modes.

To calculate the entropy, we use the geometrical model of
reference~\cite{holmes13}. In this model, the potential energy
landscape is represented as a collection of manifolds, each at
constant potential energy. The dimension of each manifold equals the
number of internal degrees of freedom of the cluster: for example, the
ground states are 0-dimensional manifolds (points), and the 8-bond
states live on 1-dimensional manifolds (lines).  To compute the
partition function, we numerically parametrize each manifold and
integrate the vibrational and rotational entropies over its entire
volume. This calculation of the entropy is purely geometrical and
requires no knowledge of the actual pair potential; the only
assumption is that the harmonic vibrational degrees of freedom
equilibrate quickly compared to motion along the soft modes.

The model reproduces our experimental measurements of the excited
state probabilities within experimental error
(Figure~\ref{fig:thryVexpt}).  The agreement validates the model's
assumption and shows that for the excited states, the entropy
associated with the soft modes dominates the permutational entropy
associated with asymmetry.  In particular, the entropy of the
zero-frequency modes explains the surprisingly high probability of
7-bond structures with 2-fold symmetry.

To understand the relative populations of the excited-state energy
levels (8-bond versus 7-bond), we must consider the interparticle
potential. Measuring the potential well is difficult because
the interaction is short-ranged---only a few tens of nanometers for the
depletion component~\cite{iracki10} and similarly ranged for the
electrostatic and van der Waals contributions.  However, the short
range makes it possible to use a ``sticky sphere''
approximation, in which a single parameter $\kappa$, called the
``sticky parameter,'' characterizes the interaction.  $\kappa$ is the
partition function for a single bond and as such is proportional to
the amount of time two particles are bound versus separated.  In the
limit where the potential becomes both infinitely narrow and
infinitely deep~\cite{holmes13},
\begin{equation} \label{eq:kappa}
\kappa = \frac{e^{-\beta U_0}}{d\sqrt {\frac{2}{\pi}\beta U''_0}}
\end{equation}
where $\beta=\frac{1}{k_BT}$, $U_0$ is the depth of the
potential well, $d$ is the microsphere diameter, and $U''_0$ is the
curvature at the potential minimum.  The advantage of this
approximation is that we need only measure $\kappa$, and not the full
potential.

We measure $\kappa$ from ratios of occupation probabilities of ground
and excited energy levels.  The total time $t_n$ for which a cluster
has $n$ bonds is proportional to $Z_n \kappa^n$, where $Z_n$ is the
sum of the partition functions of the $n$-bond
manifolds~\cite{Note1}.  By taking ratios of the time spent at
different energy levels and calculating the $Z_n$ we obtain a
measurement of the sticky parameter as $\kappa =
\frac{t_{n+1}}{t_n}\frac{Z_n}{Z_{n+1}}$.  We use observations of
smaller clusters to determine $\kappa$ independently of our 6-particle
data.  For 3-particle clusters, with 3-bond and 2-bond energy levels,
we find $\kappa$ = 29.3.  We make two more measurements of $\kappa$
using 4-particle clusters: a comparison of 5-bond to 4-bond energy
levels yields $\kappa$ = 26.8, and that of 4-bond to 3-bond levels
yields $\kappa$ = 35.3.  Using the mean of these measurements (30.5)
in the $n$-bond partition function $Z_n \kappa^n$, normalized by
$\sum_{n=7}^{9} Z_n\kappa^n$ where
$[Z_7,Z_8,Z_9]= [11900,3320,498]$, we predict 6-particle occupation
probabilities of $p_7=2.1\pm0.6\%$, $p_8=17.6\pm2.0\%$, and
$p_9=80.3\pm2.5\%$, where the uncertainties are based on the range of
measured $\kappa$ values.  The calculations agree with our
measured occupation probabilities.

The transition rates are calculated using Transition Path Theory
(TPT)~\cite{weinan10, holmes13}.  To simplify the calculations we
suppose that each transition occurs by a single bond breaking,
followed by the cluster diffusing along a 1-dimensional path and
forming another bond. We calculate the flux of probability along each
path and from this extract the non-dimensional rates, exactly as in
reference~\cite{holmes13}. The dimensional rates are obtained by
multiplying by $D/\kappa d^2$, where $D$ is the average diffusion
coefficient and $d$ is the microsphere diameter
(Table~\ref{tab:transitions}).  As our implementation of the model
ignores the time the clusters spend with fewer than 8 bonds, we expect
it to slightly overestimate the rates.

To determine the second parameter in our model, $D$, we measure the
mean-square displacements along each pathway~\cite{Note1}. The
measured diffusion coefficients range from 0.054 to 0.078 $\mu$m$^2$/s
with a mean of 0.065 $\mu$m$^2$/s (Figure~\ref{fig:transitions}). The
error bars on the measured diffusion coefficients~\cite{Note1} are
smaller than the variation in these values between the different
modes.  Thus the variation is likely due to differences in
hydrodynamic friction factors between these modes, and not measurement
error.

Nonetheless, the dimensional transition rates predicted from a simple
model using a single, average diffusion coefficient agree with the
measured rates, as shown in Table~\ref{tab:transitions}.  Using
different diffusion coefficients for each pathway yields values that
agree equally well, though not better, with the data.  This shows that
the variation in diffusion coefficients among the different modes is
not significant compared to the error in the measured transition
rates.  However, it also raises the question of why the diffusion
coefficients for different pathways vary by only about 20\% from the
mean value.  To understand this variation, we measure the diffusion
coefficient for a rearrangement in a 3-sphere cluster and find a value
of $D=$ 0.070 $\mu$m$^2$/s, close to the average value for the
6-sphere rearrangement pathways.  This agreement, along with the fact
that these diffusion coefficients are all lower than that for a single
sphere diffusing on the plane ($D=$ 0.10 $\mu$m$^2$/s), suggests that
the hydrodynamic friction factor along each pathway is dominated by
flows between those spheres that must slide or roll past one another
(as in the 3-sphere cluster), rather than by hydrodynamic interactions
between larger moving subunits of the clusters.  This would explain
why the diffusion coefficients are similar for both
diamond-square-diamond and hinge-like modes.

Taken together, these results shed new light on the free-energy
landscape, and the dynamics along it, in colloidal systems.  As in 3D
clusters, the short-range interaction in our 2D system leads to
degeneracy in both the ground and excited states.  Whereas the
occupation probabilities of the ground states are determined primarily
by symmetry (permutational entropy), those of the excited states are
determined primarily by the entropy of the soft modes.  The agreement
between the measured probabilities of the excited states and those
predicted from our geometrical model shows that the harmonic
vibrational modes equilibrate quickly compared to motion along the
soft modes.  This separation of timescales is another consequence of
the short-range interactions.  From our geometrical model of the
free-energies, we can reproduce the measured rearrangement rates
between ground states by incorporating only a single diffusion
coefficient and the partition function of a single bond, both of which
are easily measured.

Our model easily extends to 3D clusters.  Its success in describing
the 2D experimental data suggests that, at least near the isostatic
limit, it may be possible to use similar geometrically-inspired models
to understand the free-energy landscape and predict dynamics in more
complex systems with soft modes, such as bulk colloidal phases.
Indeed, such models are beginning to be developed~\cite{jenkins2014}.

\begin{acknowledgments}
  We thank Guangnan Meng, Jonathan Goodman, and David Wales for
  helpful discussions.  Rebecca W. Perry acknowledges the support of a
  National Science Foundation (NSF) Graduate Research Fellowship. This
  work was funded by the NSF through grant no. DMR-1306410 and by the
  Harvard MRSEC through grant no. DMR-0820484.
\end{acknowledgments}

\putbib
\end{bibunit}
\vspace{.5cm}
\noindent
\** \href{mailto:vnm@seas.harvard.edu}{vnm@seas.harvard.edu}


\setcounter{equation}{0}
\setcounter{figure}{0}
\setcounter{table}{0}
\onecolumngrid
\vspace{50pt}
\begin{bibunit}
\newpage

\noindent

\clearpage
\vspace*{\stretch{2}}
\thispagestyle{empty}
\begin{center}
\begin{minipage}{.6\textwidth}
\begin{center}
{\LARGE Supplemental Material}
\end{center}
\end{minipage}
\end{center}
\vspace{\stretch{3}}
\clearpage

\newpage

\section{\RNum{1}. Sample preparation protocol}
\begin{enumerate} \itemsep1pt \parskip0pt \parsep0pt
\item Prepare one small (22x22 mm) and one large (24x60 mm) glass
  coverslip (VWR Micro Cover Glasses, No. 1) by rinsing with deionized
  water, drying with high-purity compressed nitrogen, and plasma
  cleaning for 10 minutes in a PDC-32G Plasma Cleaner/Sterilizer
  (Harrick Plasma) with the RF Level set to High.\newline

\item To make a sample chamber, center the small coverslip on the
  large coverslip and separate them with narrow strips of 30 $\mu$m
  thick Mylar\textsuperscript{\textregistered} A film parallel to the
  long edges of the large coverslip.  With the two coverslips clamped
  together (e.g., with binder clips), use UV-curing Norland Optical
  Adhesive 61 and a UV lamp to seal the two edges of the small
  coverslip parallel to the spacers. We find that sealing the four
  corners and then removing the clips and sealing along the two edges
  works well. \newline

\item Use a pipette to dispense well-dispersed colloidal suspension
  near one of the unsealed edges of the small coverslip and let
  capillary action fill the sample chamber.  We use a microsphere
  volume fraction of $7.6\times10^{-6}$. \newline

\item Use Devcon 5 Minute\textsuperscript{\textregistered} Epoxy to
  seal the last two edges of the small coverslip and to go over the
  two previously sealed edges for extra protection. \newline
\end{enumerate}

\section{\RNum{2}. Image acquisition, processing, and cluster configuration identification}
To collect images, we use a Nikon Eclipse TI-E inverted microscope with a
Photon Focus camera, a CameraLink cable, and an Epix frame grabber
connected to a desktop PC. We use a combination of a 60$\times$ water
immersion objective (Nikon CFI Plan Apo VC, NA 1.2) and a 1.5$\times$
tube lens.  We choose a slow frame rate of 3 frames per second to
efficiently capture many transitions while still collecting a few
frames during each transition.  This frame rate is high enough to
allow particle tracking as described below.

By establishing four clusters of six particles in the field of view (59
$\mu$m $\times$ 59 $\mu$m), we can theoretically capture four hours of
cluster data from a typical one hour experiment.  In reality, 10 of
our 44 clusters produced data for the entire duration of the data
acquisition.  The data series from the other 34 clusters were
truncated during post-processing for one of four reasons: the cluster
diffused to the edge of the frame (7 of 44); a particle permanently
broke away from the cluster (7 of 44); the cluster came less than one
particle diameter from merging with another cluster (7 of 44); or the
particle locating or tracking algorithm failed because, for example,
the optical system drifted out of focus (13 of 44).  From 10.2 hours
of raw video, we were able to obtain 25.6 hours of 6-particle cluster
time series out of a theoretical maximum of 40.7 hours, a 63\%
recovery rate.  While we do lose track of many of our clusters over
time, this approach to data acquisition requires little supervision and
produces twice as much usable data per hour as compared to watching
over and tending to a single cluster.

Our post-processing routines are written in Python using the SciPy
ecosystem~\cite{jones01}.  We locate the particles, identify the
clusters they belong to, and track the particles from frame to
frame. To locate the particles, we first divide each image by a
background image captured with no particles in the field of view to
remove static artifacts.  We then use the Crocker and Grier
centroiding method~\cite{crocker96} to locate the particles with
better than 20 nm precision, as determined by tracking single
particles diffusing in two dimensions at 500 frames per second, and
then measuring the deviation from linearity of the mean-square
displacements at the smallest lag times.  After locating each of the
particles, we identify the cluster that each particle belongs to by
computing the distance to the four clusters' centers in the preceding
frame and selecting the cluster with the shortest distance.  We then
subtract off the cluster's center of mass from each of the particle
locations before linking them into trajectories solely using proximity
between locations in consecutive images.  Subtracting off the cluster
center of mass reduces the apparent distance moved by the particles
between frames by removing rigid-body translations.  For our
close-packed particles that occasionally diffuse distances greater
than a full particle radius between frames, subtracting off the
cluster center of mass prevents multiple particles from being linked
to a single particle in the next frame.  Alternative approaches to
tracking a collection of close-packed particles include the
optimization scheme of~\cite{crocker96} and simply using strict
proximity at a sufficiently high frame rate, where diffusing more than
a particle radius between frames is extremely unlikely.

\begin{figure}[h]
    \includegraphics[width=7cm]{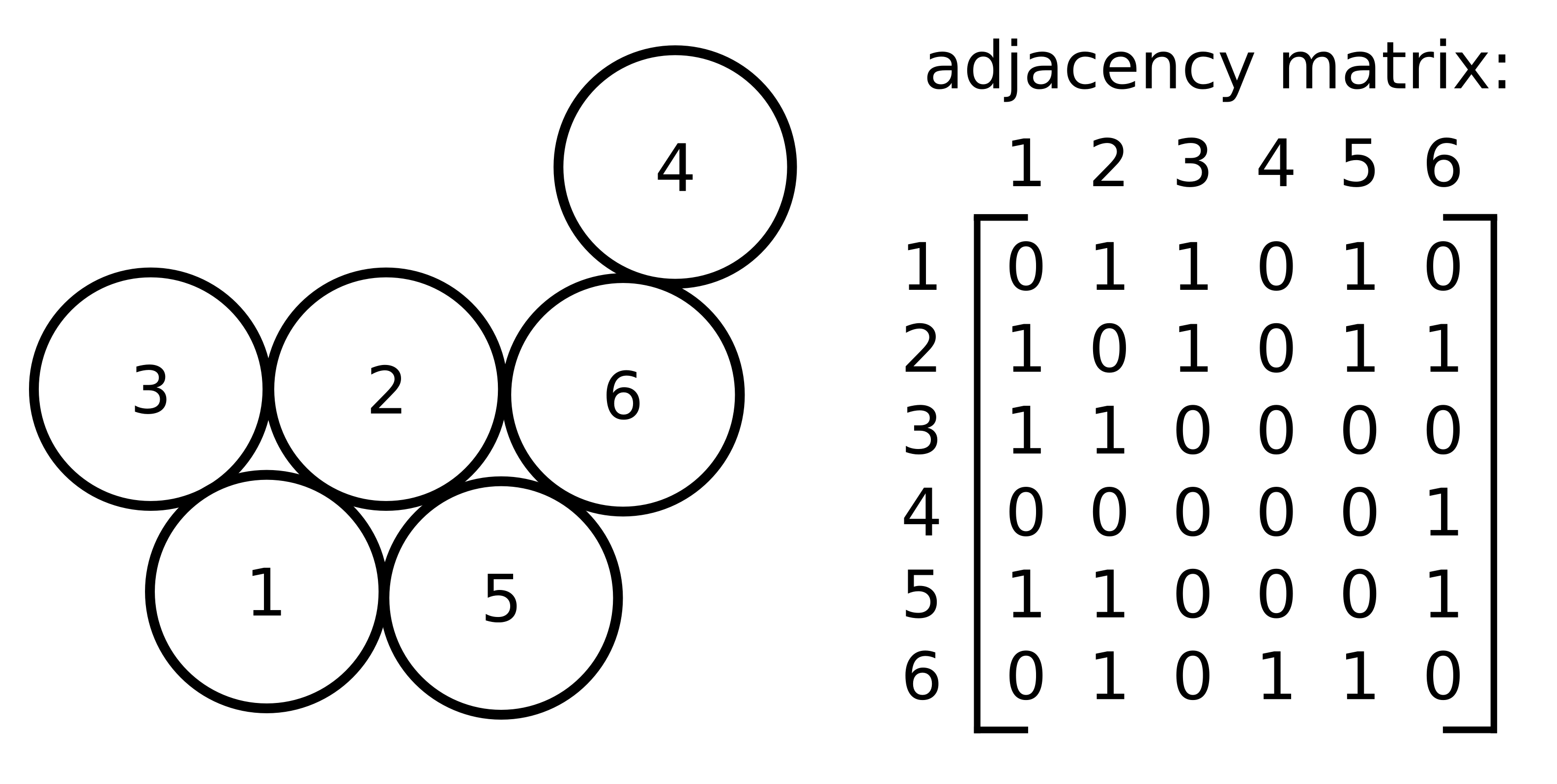}\label{adj}
    \caption{An adjacency matrix is a representation of the
      connectivity of a cluster.  Each element in the matrix relates a
      pair of particles identified by the row number and column
      number; a value of 1 signifies bound, and 0 signifies
      unbound. The adjacency matrix of the pictured 8-bond excited
      state is shown as an example.}
\label{fig:adjacency}
\end{figure}

\begin{figure}[h]
    \includegraphics[width=\textwidth]{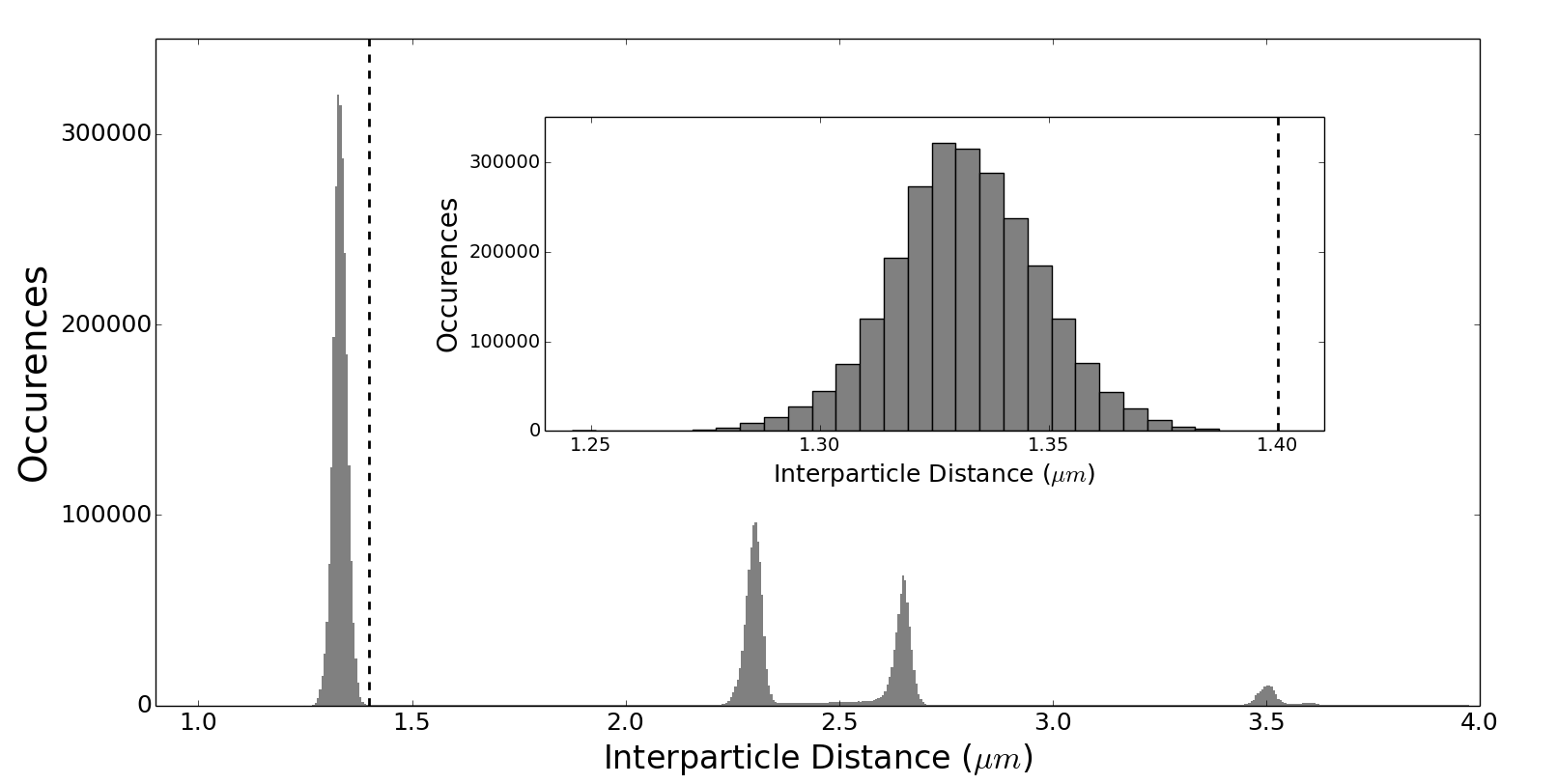}
    \caption{Distances between all particles within all 6-particle clusters at
      all times. The first peak represents bound particles at distance
      $a\approx1.33$ $\mu$m. The other peaks are at $\sqrt{3}a$, $2a$,
      and $\sqrt{7}a$ as expected for close-packed spheres on a
      plane. The width of the peaks comes from a combination of the
      particle polydispersity, the width of the interaction potential,
      and the precision of the particle locating algorithm. }
   \label{fig:cutoff}
\end{figure}

Once all the particles are found, assigned to clusters, and tracked,
we determine the configuration of each cluster in each frame by
computing the cluster's adjacency matrix~\cite{arkus_deriving_2011}
(Figure \ref{fig:adjacency}).  The adjacency matrix uniquely
determines the cluster configuration, including the particular
permutation of particles, from our library of configurations with
9-bonds, 8-bonds, 7-bonds, and ``other'' for clusters with fewer
bonds.  Such adjacency matrices do not distinguish between chiral
enantiomers, which we pair together as single configurations.  To
determine when particles are bound or unbound, we set a cutoff
distance of 1.4 $\mu$m, which is determined from the histogram in
Figure~\ref{fig:cutoff}.  We find that the occupation probabilities
are insensitive to the choice of cutoff distance.

\section{\RNum{3}. Ground state probability calculation}
Each of the macroscopic ground states---the parallelogram, chevron,
and triangle---consists of many microscopic states, so we need to
consider entropy in addition to energy in our probability
calculations~\cite{meng10}.  The probability of a macroscopic ground
state $s$ is given by the state's classical configurational integral,
$Z_s$, normalized by the sum over all the ground states:
\begin{equation}
P_s = \frac{Z_s}{\sum_{s'}{Z_{s'}}}.
\label{eqn:prob}
\end{equation}

Conveniently, $Z_s$ may be split into approximately independent
translational, rotational, and vibrational components in addition to
the contribution from the potential energy: $Z_s =
Z_{t,s}Z_{r,s}Z_{v,s}e^{-\beta U_s}$. The translational component is
identical for each ground state because the area of the glass
coverslip the clusters can explore is about seven orders of magnitude
larger than the area of a cluster.  Additionally, each of the ground
states has nine identical bonds, so the potential energy contribution is
also identical for each ground state.  By canceling out these
contributions, we arrive at a probability expression that depends only
on the rotational and vibrational components:
\begin{equation}
P_s = \frac{Z_{t,s}Z_{r,s}Z_{v,s}e^{-\beta U_s}}{\sum_{s'}{Z_{t,s'}Z_{r,s'}Z_{v,s'}e^{-\beta U_s'}}}=\frac{Z_{t}e^{-\beta U}Z_{r,s}Z_{v,s}}{Z_{t}e^{-\beta U}\sum_{s'}{Z_{r,s'}Z_{v,s'}}} =\frac{Z_{r,s}Z_{v,s}}{\sum_{s'}{Z_{r,s'}Z_{v,s'}}}.
\label{eqn:prob}
\end{equation}
The following calculations are for identical microspheres, so we
normalize the masses, interparticle distances, and spring constants to
unity.

The rotational component of the classical configurational integral in
systems of identical colloidal clusters depends on the state's moment
of inertia $I_s$, chirality $\chi_s$, and symmetry number $\sigma_s$,
which accounts for the effects of
permutations~\cite{gilson_symmetry_2010}:
\begin{equation}
Z_{r,s} \propto \frac{\chi_s\sqrt{I_{s}}}{\sigma_s}.
\end{equation}  
The moment of inertia is more generally the determinant of the moment
of inertia tensor, but here the cluster has only one rotational axis.
The chirality $\chi_s$ is 1 if the configuration is achiral and 2 if
the configuration is a pair of chiral enantiomers.

To compute the vibrational contribution to the ground state
probabilities, we use the harmonic approximation for the interparticle
interactions.  The vibrational contribution is inversely proportional
to the product of the frequencies of the normal modes.  There are
$2N-3$ normal modes, since there are $2N$ degrees of freedom, and we
have already removed 2 translational degrees of freedom and accounted
for 1 rotational degree of freedom.  The vibrational frequencies are
given by the square root of the non-zero eigenvalues of the matrix
$\boldsymbol{H_s}$, constructed from $N\times N$ super-elements. Each
super-element is a $2\times2$ Hessian matrix describing the
interactions between particles $i$ and $j$~\cite{eyal06}:
\begin{equation}
H_{ij} = \begin{bmatrix} \frac{\partial^2U}{\partial x_i \partial x_j} & \frac{\partial^2U}{\partial x_i \partial y_j} \\ \frac{\partial^2U}{\partial x_j \partial y_i} & \frac{\partial^2U}{\partial y_i \partial y_j} \end{bmatrix}.
\end{equation}

The eigenvalues $k_{\alpha,s}$ of $\boldsymbol{H_s}$ are the squares of
the normal mode frequencies, which allow us to compute the vibrational
contribution to the classical configuration integral:
\begin{equation}
Z_{v,s} \propto \prod_{\alpha = 1}^{2N-3}{ \sqrt{\frac{1}{k_{\alpha,s}}}}.
\end{equation}
This expression for the vibrational contribution is the last piece we
need in order to use Equation~\ref{eqn:prob} to calculate the
probabilities of the parallelogram, chevron, and triangle.  The
results are presented in Table~\ref{tab:components}.

 \begin{table}

 \begin{ruledtabular}

	\begin{tabular}{cccccccc}
	&&  & Parallelogram & Chevron & Triangle&&\\
\noalign{\vskip 2mm}  
	&&$\sqrt{I_s}$ & $\sqrt{5\frac{1}{2}}$ & $\sqrt{4 \frac{5}{6}}$ & $\sqrt{5}$ &&\\
	&&$\chi_s$ & 2 & 1 & 1 &&\\
	&&$\sigma_s$ & 2 & 1 & 3 &&\\
	&&$Z_{r,s}$ &  $\sqrt{5\frac{1}{2}}$ & $\sqrt{4 \frac{5}{6}}$ & $\frac{1}{3}\sqrt{5}$ &&\\
	&&$Z_{v,s}$ & $\frac{8}{27}\sqrt{\frac{2}{11}}$ & $\frac{8}{27}\sqrt{\frac{6}{29}}$ & $\frac{8}{27}\sqrt{\frac{1}{5}}$ &&\\
\noalign{\vskip 2mm}  
\hline
\noalign{\vskip 2mm}  
	&&Probability & $\frac{3}{7}$ & $\frac{3}{7}$ & $\frac{1}{7}$ &&\\
\noalign{\vskip 2mm}  

	\end{tabular}

 \end{ruledtabular}
\caption{Comparison of the components factoring into the probabilities
  of the 3 ground states for two-dimensional clusters of 6 particles.} 
\label{tab:components}
 \end{table}

\section{\RNum{4}. Occupation probability error bars}

The empirical occupation probability of each excited state is computed
by taking the total amount of time we observe its adjacency matrix,
and dividing by the total time spent in all configurations with
identical energy.  To estimate the error bar on this statistic we need
to know the number of effectively independent samples. In general this
is not the same as the number of data points, since the data are
correlated in time: if a cluster has a particular adjacency matrix
during one time step, it it more likely to remain in that adjacency
matrix in subsequent time steps. After enough time steps, however, the
data becomes decorrelated, and only then can new data be treated as
independent.  Roughly, the number of effectively independent samples
is the length of the data, divided by the ``correlation time'' of the
data.

A cluster can be thought of as a stochastic process $X_t \in
\mathbb{R}^{2N}$, where $X_t$ lists the positions of the particles. An
adjacency matrix corresponds to a subset $A\subset\mathbb{R}^{2N}$ of
configuration space. We would like to know the average amount of time
the system spends in set $A$, which we write as $p_A = \E 1_{(X_t\in
  A)}$.

Let's define a process $X_A(t) \equiv 1_{(X_t\in A)}$ to be the
process that is 1 if $X(t)\in A$, and 0 otherwise. Then $p_A = \E
X_A(t)$.  Let $\hat{p}_A = \oneover{T}\int_0^T X_A(t)dt$ be an
estimator for $p_A$.  Let's suppose this estimator is Gaussian, i.e.
$\hat{p}_A = p_A + \sigma_A z_A$, where $\sigma_A$ is the standard
deviation of the estimator, and $z_A\sim N(0,1)$ is a copy of the
standard normal.  Then, we can construct construct 95\% error bars as
$e = 1.96\sigma_A$.

How can we determine the standard deviation
$\sigma_A$?  If each observation were independent, then we would have
$
\sigma_A^2 = \frac{\sigma_{A,0}^2}{n},
$
where $\sigma_{A,0}$ is the standard deviation of $X_A(t)$ at a single
point in time (equal to $p_A(1-p_A)$ for our process since it's an
indicator function), and $n$ is the number of independent
observations.

For a process that is correlated in time, a similar result holds
provided we replace $n$ with the number of ``effectively'' independent
samples~\cite{sokal}. This is given by $n_{\mbox{\em \scriptsize eff}}
= T/\tau$, where $T$ is the total length of time of the sample, and
$\tau$ is the correlation time. The correlation time is defined (for a
stationary process) from the correlation function $C_A(t) \equiv \E
X_A(s)X_A(s+t)$ to be
\begin{equation}
\tau = \oneover{C_A(0)}\int_{-\infty}^\infty C(t)dt. 
\end{equation}
Geometrically, this comes from taking all the area under the
correlation function and forming it into a rectangle with the same
height as the covariance function at $t=0$, so the width is
$\tau$. Note that $C_A(0) = \sigma_{A,0}^2$.

The estimate for $\sigma_A^2$ is then
\begin{equation}\label{eq:sigma}
 \hat{\sigma}_A^2 = \frac{\sigma_{A,0}^2}{n_{\mbox{\em \scriptsize eff}}} = \oneover{T}\int_{-\infty}^\infty C_A(t)dt.
\end{equation}
We have used the fact that $\sigma_{A,0} = C_A(0)$ to rewrite the
integral.  This integral is calculated numerically from the data
following the algorithm described in section B.

\subsection{A. Conditional probabilities}

The numbers we report in manuscript Figures 1 and 2 are conditional
probabilities: the probability of the cluster having a particular
adjacency matrix, conditional on it having a given number of bonds. 
Calculating the variance of these conditional
probabilities requires extra considerations.

Suppose we want to estimate the relative probability of being in set
$A$, conditional on also being in a set $B$. That is, we would like to
estimate $p_{A|B} = P(X(t)\in A | X(t) \in B)$ = $\frac{P(X(t)\in
  A)}{P(X(t)\in B)} = \frac{\E 1_{(X(t)\in A)}}{\E 1_{(X(t)\in
    B)}}$. Let $X_B(t) = 1_{(X(t)\in B)}$. Then an estimator for
$p_{A|B}$ is $\hat{p}_{A|B} = \frac{\hat{p}_A}{\hat{p}_B}$. When
$\sigma_i$ is small, this can be expanded as:
\begin{equation*}
\frac{\hat{p}_A}{\hat{p}_B} = \frac{p_A + \sigma_A z_A}{p_B + \sigma_B z_B} 
  = \frac{p_A}{p_B} + \frac{\sigma_Az_A}{p_B} - \frac{p_A\sigma_Bz_B}{p_B^2} + O(\sigma_i^2).
\end{equation*}
The variance of this estimator for small $\sigma_i$ is approximately 
\begin{equation}
\text{var}\left(\frac{\hat{p}_A}{\hat{p}_B}\right) = \frac{\sigma_A^2}{p_B^2} + \frac{p_A^2\sigma_B^2}{p_B^2} - \frac{2p_A\sigma_A\sigma_B\E z_Az_B}{p_B^2} 
= \frac{\sigma_A^2}{p_B^2} + \frac{p_A^2\sigma_B^2}{p_B^2} - \frac{2p_A\sigma^2_{AB}}{p_B^2}.
\end{equation}
We can estimate $\sigma_A, \sigma_B$ as in the previous section. To
compute the cross-correlation term $\sigma_A\sigma_B\E z_Az_B =
\sigma_{AB}^2$, we compute the cross-correlation function $C_{AB}(t) =
\E X_A(s)X_B(s+t)$ and determine the variance from this, as in the
previous section.

\subsection{B. How to compute the correlation time $\tau$}

The correlation function is very noisy at late times, so the integral
to compute $\tau$ will also be very noisy. In fact, the bias as
$n\to\infty$ is 0, but the variance is $O(1)$. Therefore that integral
is not a consistent estimator of $\tau$~\cite{sokal}.

We use a windowing method to estimate $\tau$, which integrates the
correlation function up to a multiple $W$ of the current estimate of
$\tau$. As is commonly done we set $W=5$.  Here is the method in
pseudo-code:
\begin{quote}
\texttt{
\hspace{-0.2cm}$\hat{\rho}_t = C(t)/C(0)$\\ 
$\tau=1$\\
$t=1$\\
while($\tau<Wt$) 
\{\\
\qquad $\tau = \tau + 2\hat{\rho}_t$\\
\qquad $t=t+1$\\
\}
}
\end{quote}
This produces an estimator whose variance goes to zero as the number
of samples increases, but with a small bias of size $O(e^{-W})$ (if
the covariance function has exponential tails.)

\subsection{C. Why this works}
Here is a brief explanation for Equation \eqref{eq:sigma}. The variance of
$\hat{p}_A$ is
\begin{align*}
\left(\oneover{T}\int_0^TX_A(t)dt \right)\left(\oneover{T}\int_0^TX_A(s)ds \right) - p_A^2 &= \oneover{T^2}\int_0^T\int_0^TC_A(t-s)dt\,ds\\
 &= \oneover{2T^2}\int_{-T}^T\int_{u}^{2T-u} C_A(u)dv\,du\\
 &= \oneover{T}\int_{-T}^T C_A(u)(1-\frac{u}{T})du\\
 &\approx \oneover{T}\int_{-\infty}^\infty C_A(u)du.
\end{align*}
The last approximation is valid when $T$ is large enough that $C_A(u)$
has decayed.

\section{\RNum{5}. Measuring the diffusion coefficients}

\begin{figure}[h]
    \includegraphics[width=7cm]{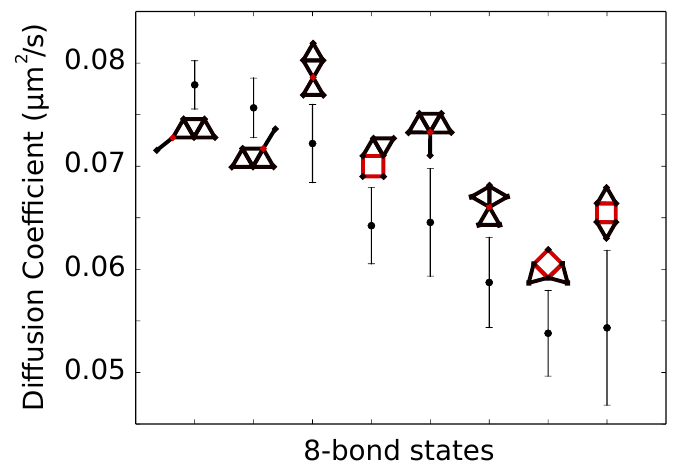}\label{diff}
    \caption{Measured diffusion coefficients for the one-dimensional soft modes of the 8-bond states.  Hinge-like
        joints and non-rigid squares are labeled in red.  The error bars are 95\% confidence intervals.}
\label{fig:diff}
\end{figure}

To measure diffusion coefficients, we must first parameterize each of
the one-dimensional transition paths between rigid clusters. A cluster
can be written as a vector $x\in\mathbb{R}^{2N}$ 
 listing the centers of each sphere in two dimensions. We
find a path $x(s)$ depending on parameter $s$, such that
\begin{enumerate*}
\item $\dd{x}{s}$ is perpendicular to infinitesimal rotations,
  infinitesimal translations, and motions that change the bond lengths

\item $|\dd{x}{s}| = 1$.
\end{enumerate*}
The first is possible because the space of rotations, translations,
and bond lengths is ($2N-1$)-dimensional 
 since there is exactly one bond ``missing,'' so at each
point along the path there is a one-dimensional tangent space spanned
by unit vector $t_s$. The second is possible because the space we are
parameterizing is one-dimensional, so we can always find an arc-length
parameterization.

We store the path as a discrete set of clusters $x_{s_0}, x_{s_1},
\ldots, x_{s_m}$, where $s_k= k\Delta s$ for fixed step size $\Delta
s$. Each $x_{s_{i+1}}$ is found from $x_{s_i}$ by taking a step of
size $\Delta s$ in the direction of the unit tangent $t_{s_i}$, and
then orthogonally projecting back to the manifold of constraints:
$x_{s_{i+1}} = P(x_{s_i} + t_{s_i}\Delta s)$, where $P$ is an
orthogonal projection operator. The details of $P$ are provided in
reference~\cite{holmes13}.

We next analyze our data to obtain a time series of $s$-values along
each transition path. For each data point with 8 bonds, we find its
corresponding $s$-value by first performing an orthogonal projection
onto the transition path to remove the vibrational degrees of
freedom. This projection step was crucial to obtaining good
statistics. Then, we identify the closest cluster in the list
$\{x_{s_0}, \ldots, x_{s_m}\}$, using a Euclidean metric on the space
of sorted bond distances. Finally, for each pair of consecutive points
that lie on the same transition path with $s$-values
$\hat{s}_1,\hat{s}_2$, we compute the change in $s$-values $\Delta =
\hat{s}_2- \hat{s}_1$.

The result is a sequence of increments $\Delta_1 ,\Delta_2,\ldots,
\Delta_M$ associated with each
transition path.  Close to the ends of the manifolds, the allowed
sizes of steps taken towards the end become more and more restricted
by the end of the manifold.  To avoid biasing due to the non-Gaussian
distributions of $\Delta$ near the ends of the manifolds, we only
analyze steps towards the center of the manifold.  Since the velocity
correlation time is much shorter than the time between measurements,
the cluster performs Brownian motion along the transition path, so the
average diffusion coefficient along a path can be estimated as $D =
a^2\oneover{2\Delta t}\oneover{M}\sum_{i=1}^M (\Delta_i^2)$. Here
$\Delta t$ is the time between measurements, and the average is with
respect to the stationary distribution along each path.  The square of
the interparticle spacing, $a^2$, is the conversion factor between
diffusion in the parameterized space and in real-space.  The values we
arrive at are between 0.05 and 0.08 $\mu$m$^2$/s as shown in
Figure~\ref{fig:diff}.

\section{\RNum{6}. Table of $Z_n$ for clusters with $N\leq$ 6}

To compute the sticky parameter, $\kappa$, we need to know the total
geometrical partition function, $Z_n$, for manifolds with $n$ bonds, for
at least two different values of $n$. The ``geometrical'' partition
function is the part which comes from integrating the rotational and
vibrational partition functions; this is geometrical because it depends only on
the locations, shapes, and sizes of the particles, and not on the
potential energy or temperature.

The total geometrical partition function is 
\begin{equation}
Z_n = \sum_i z^{(n)}_i, 
\end{equation}
where $z^{(n)}_i$ is the geometrical partition function for a single
manifold with $n$ bonds, and the index $i$ runs over all manifolds
with $n$ bonds. The geometrical partition function for a single
manifold $\Omega^{(n)}_i$ is
\begin{equation}\label{eq:zi}
z^{(n)}_i = \int_{\Omega^{(n)}_i} h^{(n)}_i(y)I^{(n)}_i(y)d\sigma_{\Omega^{(n)}_i}(y),
\end{equation}
where $d\sigma_{\Omega^{(n)}_i}(y)$ is the volume element on the
manifold, $I^{(n)}_i(y)$ is the rotational partition function, and
$h^{(n)}_i(y)$ is the ``geometrical'' part of the vibrational
partition function. The latter equals $\prod_j\lambda_j^{-1/2}$, where
$\lambda_j$ are the non-zero eigenvalues of the Hessian of the
potential energy, in the harmonic approximation with the spring
constant set to 1.

To compute~Equation \eqref{eq:zi} numerically, we parameterize each
manifold and use a finite-element method to compute the
integral. The supplemental information of reference
\cite{holmes13} contains more details on how to compute the
parameterization and volume element.

Table \ref{tab:zs} lists the numerically computed values of the total
geometric partition function for the 0, 1, and 2-dimensional
manifolds.

\begin{table}[h!] 

\caption{The following geometrical partition functions are generated
  by applying the methods from reference~\cite{holmes13} to 2D clusters. Note:
  clusters with a single disconnected sphere are not included in these
  calculations.}


\begin{ruledtabular}
	\begin{tabular}{cccccccc}
	&&N & $Z_{2N-3}$ & $Z_{2N-4}$ & $Z_{2N-5}$&&\\
	&&3 & 0.770 & 4.19 & -- &&\\
	&&4 & 4.00 & 23.4 & 60.2 &&\\
	&&5 & 37.0 & 231 & 763 &&\\
	&&6 & 498 & 3320 & 11900 &&\\
	\end{tabular}\label{tab:zs}
\end{ruledtabular}

\label{tab:geometricalcont}
\end{table}

\section{\RNum{7}.  Realtime\_transitions.avi}

Video segments show the 8-bond transitions between ground states.  The
clusters transition from the ground state pictured on the left to the
ground state pictured above.  Connectivity diagrams label the excited
state shown in each movie.  The micrographs were divided by a
background to remove static artifacts and scaled to create identical
background intensities. We created this compilation using the
Matplotlib library~\cite{hunter07}. Videos are played back at the
recording rate of 3 frames per second.

\section{\RNum{8}.  10xfast\_fourClusters.avi}
This clip of 11 minutes (2000 frames) of raw data shows our
experimental arrangement for simultaneously observing 4 clusters of 6
spheres while they rotate, translate, and rearrange. The clusters
rearrange frequently, but rarely break apart. Playback is 10 times faster
than recorded.


\putbib

\vspace{.5cm}
\noindent
\** \href{mailto:vnm@seas.harvard.edu}{vnm@seas.harvard.edu}

\end{bibunit}

\end{document}